\documentclass[aps,letterpaper,twocolumn,prl,showpacs]{revtex4-1}
\usepackage{amsmath,bm}
\usepackage{graphicx,color}
\usepackage[english]{babel}
\usepackage{hyperref}

\begin{document}

\title{Writing and Erasing of Temporal Kerr Cavity Solitons via Intensity Modulation of the Cavity Driving Field}

\author{Yadong Wang$^{1}$}
\email{ywan505@aucklanduni.ac.nz}
\author{Bruno Garbin$^{1}$}
\author{Fran\c{c}ois Leo$^{2}$}
\author{St\'ephane Coen$^{1}$}
\author{Miro Erkintalo$^{1}$}
\email{m.erkintalo@auckland.ac.nz}
\author{Stuart G. Murdoch$^{1}$}

\affiliation{$^{1}$The Dodd-Walls Centre for Photonic and Quantum Technologies, Department of Physics, The University of Auckland, Auckland 1142, New Zealand \\
$^{2}$OPERA-photonics, Universit\'e libre de Bruxelles, 50 Avenue F. D. Roosevelt, CP 194/5, B-1050 Bruxelles, Belgium
}

\begin{abstract}
We experimentally and numerically study the use of intensity modulation for the controlled addressing of temporal Kerr cavity solitons. Using a coherently driven fiber ring resonator, we demonstrate that a single temporally broad intensity modulation pulse applied on the cavity driving field permits systematic and efficient writing and erasing of ultrashort cavity solitons. We use numerical simulations based on the mean-field Lugiato-Lefever model to investigate the addressing dynamics, and present a simple physical description of the underlying physics.

\end{abstract}

\maketitle

\noindent Temporal Kerr cavity solitons (CSs) are pulses of light that can circulate indefinitely in coherently-driven passive nonlinear cavities \cite{wabnitz_suppression_1993, leo_temporal_2010}. They exist through a delicate double balance: self-focussing Kerr nonlinearity compensates for anomalous group-velocity dispersion, while cavity losses are compensated by nonlinear interactions with the continuous wave (cw) beam driving the cavity. These temporal CSs display behaviours analogous to those of spatial CSs that have been extensively studied in the context of \emph{diffractive} nonlinear cavities~\cite{barland_cavity_2002, ackemann_chapter_2009}. In particular, multiple CSs can simultaneously co-exist, and they can be individually addressed, i.e., turned on (writing) or off (erasure).

Temporal CSs were first observed in 2010 in a passive fiber ring resonator \cite{leo_temporal_2010}. Their dynamics and characteristics have subsequently been investigated using similar fiber-based systems~\cite{jang_ultraweak_2013, luo_spontaneous_2015, anderson_observations_2016}. More recently, temporal CSs have also been observed in high-Q optical microresonators \cite{herr_temporal_2014, Yi_soliton_2015, joshi_thermally_2016, webb_experimental_2016, obrzud_temporal_2017}, where they have been shown to underpin the formation of coherent optical frequency combs. These findings have attracted significant interest in CS-related research, thanks to the many  important application prospects of microresonator frequency combs~\cite{griffith_silicon-chip_2015, suh_microresonator_2016, pasquazi_micro-combs:_2017, marin-palomo_microresonator-based_2017}.

Excitation of CSs requires that the intracavity cw field is suitably perturbed~\cite{jang_writing_2015}. Various techniques have been reported in the literature: CSs have been written (i) by adiabatically changing the system parameters over several cavity round trips (e.g. by scanning the pump-cavity detuning or ``power kicking'' the driving laser)~\cite{luo_spontaneous_2015, anderson_observations_2016, herr_temporal_2014, Yi_soliton_2015, joshi_thermally_2016, webb_experimental_2016, obrzud_temporal_2017}; (ii) through cross-phase modulation (XPM) induced by pulses from an external mode-locked laser~\cite{leo_temporal_2010}; and (iii) via direct phase modulation of the cavity driving field~\cite{jang_writing_2015}. Each of these techniques has its own advantages and disadvantages. Adjustment of the system parameters is the most straightforward technique to implement, but does not provide full control over the soliton configuration. The XPM method allows for reliable writing of arbitrary CS patterns, but requires a mode-locked laser which increases the system's complexity. Direct phase modulation reduces that complexity, and is the only method that has been experimentally demonstrated to permit selective erasure of temporal CSs. Unfortunately, several consecutive phase modulation pulses (synchronised to the cavity roundtrip time) are needed to achieve successful writing and erasing~\cite{jang_writing_2015}, limiting the system's speed and efficiency. McDonald and Firth have theoretically shown (in the context of spatial CSs) that direct \emph{amplitude} modulation applied over a single cavity transit could allow for more efficient addressing~\cite{mcdonald_all-optical_1990, mcdonald_switching_1993}. More recent theoretical works have also suggested amplitude modulation to represent a viable route for the controlled excitation of CSs in microresonators~\cite{kang_deterministic_2017}. Yet, no experimental demonstrations have hitherto been reported.

\looseness=-1In this Letter, we demonstrate and study the controlled addressing of temporal CSs via intensity modulation of the cavity driving field. Specifically, we show that both writing and erasing can be achieved by applying a single intensity modulation pulse directly on the cavity driving field. This method possesses all the advantages of the phase modulation technique, but additionally relaxes the need for several addressing pulses synchronised to the cavity roundtrip time. We also present a simple interpretation of CS excitation via intensity perturbations.

We first present results from illustrative numerical simulations. To this end, we model the nonlinear cavity dynamics using the mean-field Lugiato-Lefever equation (LLE)~\cite{haelterman_dissipative_1992, coen_modeling_2013}:
\begin{equation}
\begin{split}
\label{eq:LLE}
t_{\mathrm{R}} \frac{\partial E(t,\tau)}{\partial t} =& \left(-\alpha + i \gamma L |E|^2 - i \delta_0 - i \frac{\beta_2 L}{2} \frac{\partial^2}{\partial \tau^2} \right)E \\ &+ \sqrt{\theta P_\mathrm{in}(t,\tau)}.
\end{split}
\end{equation}
\noindent Here $t$ is the slow time describing the evolution of the intracavity electric field envelope $E(t,\tau)$ over successive round trips, while $\tau$ is a fast time describing the envelope's temporal profile. $t_\mathrm{R}$ is the round trip time, $\alpha$ is equal to half the fraction of power lost per round-trip, $\gamma$ is the Kerr nonlinearity coefficient, $L$ is the length of the resonator, $\delta_0$ is the phase detuning of the driving laser from the closest cavity resonance, and $\beta_2$ is the group-velocity dispersion coefficient. Finally, $\theta$ is the power transmission coefficient of the coupler used to inject the driving field, with power profile $P_\mathrm{in}(t,\tau)$, into the cavity.

\begin{figure}[t]
	\centering
	\includegraphics[width=\linewidth]{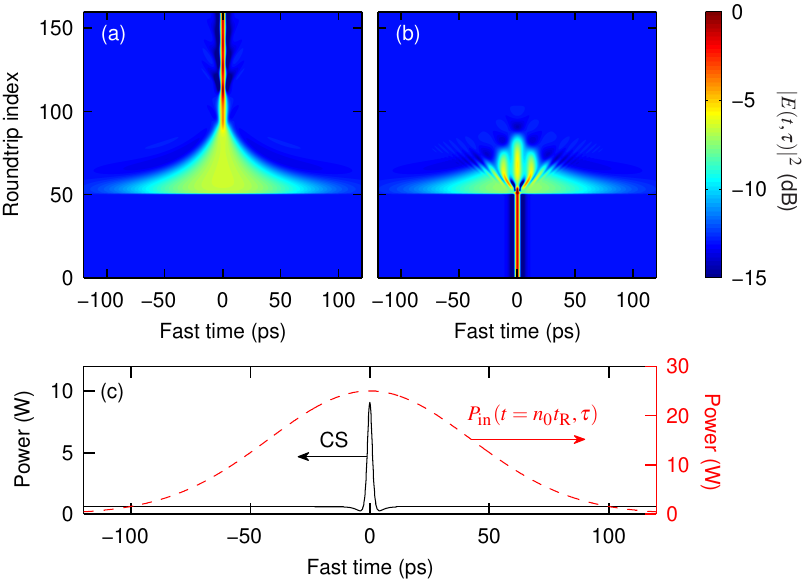}
	\caption{\small Numerical simulation results showing the (a) writing and (b) erasing of a temporal CS with a single intensity modulation pulse. As shown in (c), the intensity modulation pulse is almost two orders of magnitude broader than the excited CS. The simulation parameters are: $\alpha = 0.15$, $\delta_0 = 0.45~\mathrm{rad}$, $\gamma = 1.2~\mathrm{W^{-1}}\mathrm{km^{-1}}$, $L = 100~\mathrm{m}$, $\beta_2 = -21.4~\mathrm{ps^2km^{-1}}$, $\theta = 0.1$, $P_\mathrm{cw} = 1~\mathrm{W}$. In (a) and (b) the peak power of the addressing pulse is $P_\mathrm{p} = 18.5~\mathrm{W}$ and $P_\mathrm{p} = 15~\mathrm{W}$, respectively, and $n_0 = 50$. The logarithmic colormap is normalized to the maximum of the individual simulations.}
	\label{fig1}
\vskip-10pt
\end{figure}

In the experiments that follow, we realise CS writing and erasing by applying a single intensity modulation pulse on top of an otherwise cw driving field. Assuming  a Gaussian addressing pulse, we model the driving field power profile as
\begin{equation}
P_\mathrm{in}(t,\tau) = P_\mathrm{cw} + P_\mathrm{p}e^{-\tau^2/\Delta\tau^2}f(t),
\end{equation}
where $P_\mathrm{cw}$ and $P_\mathrm{p}$ describe the power of the cw driving field and the peak power of the addressing pulse, respectively, and $\Delta\tau$ is the width of the intensity modulation pulse. $f(t) = \Pi(t/t_\mathrm{R}-n_0-1/2)$ is a rectangle function that is unity for $t\in [n_0t_\mathrm{R}, (n_0+1)t_\mathrm{R})$ and zero elsewhere, thus modelling a gate that permits a single intensity modulation pulse to enter the cavity at the beginning of round trip number $n_0$. It is worth emphasizing that, in contrast to recent microresonator configurations that are continuously driven with short optical pulses~\cite{obrzud_temporal_2017}, our system is predominantly cw-driven; the intensity modulation pulses are only used for writing and erasing. We also note that, whilst the use of a single excitation pulse may appear to violate the tenets behind the mean-field Eq.~\eqref{eq:LLE}, we have carefully verified that a more realistic model based on an Ikeda-like cavity map yields similar results.

Figure~\ref{fig1}(a) shows numerically simulated writing dynamics for parameters similar to the experiments that will follow  (see caption). The simulation initial condition corresponds to the stable cw solution of the system. At round trip $n_0 = 50$, a single intensity modulation pulse with a full-width at half maximum (FWHM) of $1.66\Delta\tau = 100~\mathrm{ps}$ is applied on the driving field. As can be seen, the intensity modulation reshapes into a temporal CS in about 50 roundtrips. To demonstrate erasure, we perform another simulation with an initial condition corresponding to a single temporal CS centered at $\tau = 0$. To avoid re-excitation, a smaller addressing pulse peak power is used (see caption). Results are shown in Fig.~\ref{fig1}(b), and we readily see how the intensity modulation pulse (applied at round trip $n_0 = 50$) erases the CS, bringing the intracavity field to the cw state. At this point, we highlight that 100~ps FWHM intensity modulation pulses can be readily realized using standard 10~GHz intensity modulators. Yet, as our simulations show, they lead to the generation of 2.6~ps CSs --- two orders of magnitude shorter.

\begin{figure}[t]
	\centering
	\includegraphics[width=\linewidth]{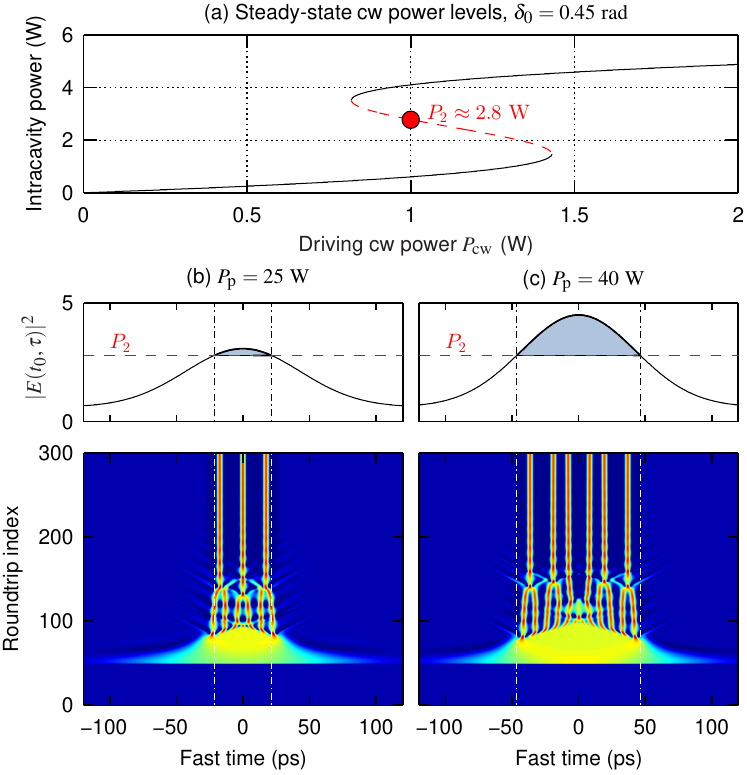}
	\caption{\small (a) cw steady-state solutions of the LLE with parameters as in Fig.~\ref{fig1}. (b, c) CS addressing dynamics for peak powers (b) $P_\mathrm{p} = 25~\mathrm{W}$ and (c) $P_\mathrm{p} = 40~\mathrm{W}$. The top panels in (b, c) show intracavity field profiles at $t_0 = n_0t_\mathrm{R}$, i.e., immediately after the intensity modulation pulse is launched into the cavity. The shaded areas highlight regions above the critical power $P_2$. Colormap as in Fig.~\ref{fig1}(a, b).}
	\label{fig2}
\vskip-10pt
\end{figure}

\looseness=-1CS writing and erasure is, in general, a complex dynamical process that depends non-trivially on several parameters~\cite{jang_writing_2015, mcdonald_switching_1993}, with only certain parameter configurations resulting in successful addressing. In the context of amplitude perturbations, we expectedly find that successful writing and erasing is achieved only when the amplitude of the addressing pulse is above a certain threshold (which depends on the cavity detuning, driving power, addressing pulse width, ...). On the other hand, if the amplitude is too large,  several CSs can inadvertently emerge. This latter observation is amenable to a simple qualitative interpretation. Specifically, CS excitation can be expected when the peak power of the intracavity field after application of the perturbation is sufficiently close (or exceeds) the critical power of the unstable cw solution of the cavity system. In this case, the intensity modulation may switch the lower state cw solution locally to the upper state, resulting in the formation of CSs via modulation instability (MI). A single CS can be expected if the width of the intracavity region above the critical power is shorter (or similar) than the characteristic MI time scale, whilst several can emerge in the opposite limit.

We illustrate this interpretation in Fig.~\ref{fig2}. Figure~\ref{fig2}(a) first shows the power levels of the cw steady-state solutions for a fixed detuning $\delta_0 = 0.45~\mathrm{rad}$ as a function of the driving cw power. For $P_\mathrm{cw} = 1~\mathrm{W}$ considered in the simulations that will follow, the power level of the unstable cw state is $P_2 \approx 2.8~\mathrm{W}$. In Figs.~\ref{fig2}(b) and (c), we show results from two simulations with different perturbation peak powers $P_\mathrm{p} = 25~\mathrm{W}$ and $P_\mathrm{p} = 40~\mathrm{W}$, respectively [other parameters as in Fig.~\ref{fig1}]. The top panels show the intracavity field in the beginning of round trip 51, immediately after the injection of the intensity modulation pulse. In both cases, we see that a region [shaded in blue in Fig.~\ref{fig2}(b,c)] of the intracavity field lies above the critical power $P_2 \approx 2.8~\mathrm{W}$, and how a pattern of CSs forms within that region. The peaks in the nascent pattern are initially spaced by about 8.5~ps, in excellent agreement with the calculated MI period of about 8.9~ps. Following the initial break-up into a quasi-periodic pattern, the CSs collide and merge~\cite{jang_controlled_2016} with one another, but eventually reach a steady-state configuration consisting of several CSs.

\begin{figure}[t]
	\centering
	\includegraphics[width=\linewidth, clip = true]{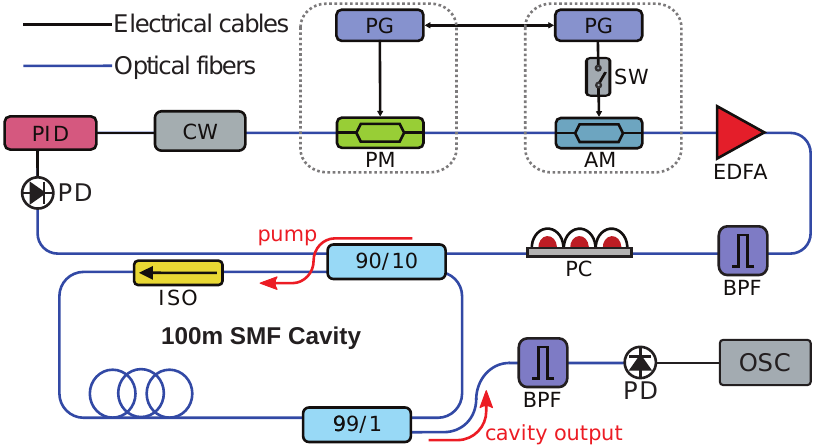}
	\caption{\small Schematic of the experimental setup. PD, photodetector; PID, proportional-integral-derivative; PG, pattern generator; SW, switch; PM, phase modulator; AM, amplitude modulator; EDFA, Erbium-doped fiber amplifier; BPF, band-pass filter; PC, polarization controller; ISO, isolator; OSC, oscilloscope; SMF, single-mode fiber.}
	\label{fig3}
\vskip-10pt
\end{figure}

We now proceed to describe our experimental demonstrations of CS writing and erasing using intensity modulation. Our experiments were performed using the setup shown in Fig. \ref{fig3}, which overall is similar to that in~\cite{jang_writing_2015}. The experiment is built around a fiber ring resonator made out of 100~m of single mode optical fiber that is closed on itself with a 90/10 coupler. An optical isolator is used to suppress stimulated Brillouin scattering and a 99/1 tap-coupler allows for the intracavity dynamics to be monitored. To improve the signal-to-noise ratio, an off-centered (1551.2 nm) band-pass filter with a 0.6 nm bandwidth is used to remove the cw field component before detection. The cavity finesse ($\mathcal{F} = \pi/\alpha$) was measured to be 21. The cavity is driven with a 1550~nm narrow-linewidth cw laser that is amplified to about 1~W. An electronic servo system is used to actively lock the reflected signal from the 90/10 coupler to a set level, thus stabilizing the cavity detuning~\cite{jang_ultraweak_2013}. 

The key component that performs the writing and erasing in the setup is the 10 GHz optical amplitude modulator. It imprints electrical pulses, derived from an electrical pattern generator, onto the cw driving field. An electronic switch is used to gate the periodic output of the pattern generator, ensuring that only a single addressing pulse is launched into the cavity. We measured the duration of the optical addressing pulses to be 120~ps just before they are launched into the resonator. To demonstrate erasure, the intensity modulation pulses are synchronised with the solitons in the cavity by imparting a 10~GHz sinusoidal phase modulation on the driving beam before the amplitude modulator; the ensuing intracavity phase modulation robustly traps the CS to the modulation maxima~\cite{jang_temporal_2015}, thus providing for precise timing references. Erasure can then be achieved by carefully aligning the phase and amplitude modulation profiles such that they overlap and are synchronized to the cavity round trip time.

\begin{figure}[t]
	\centering
	\includegraphics[width=\linewidth, clip = true]{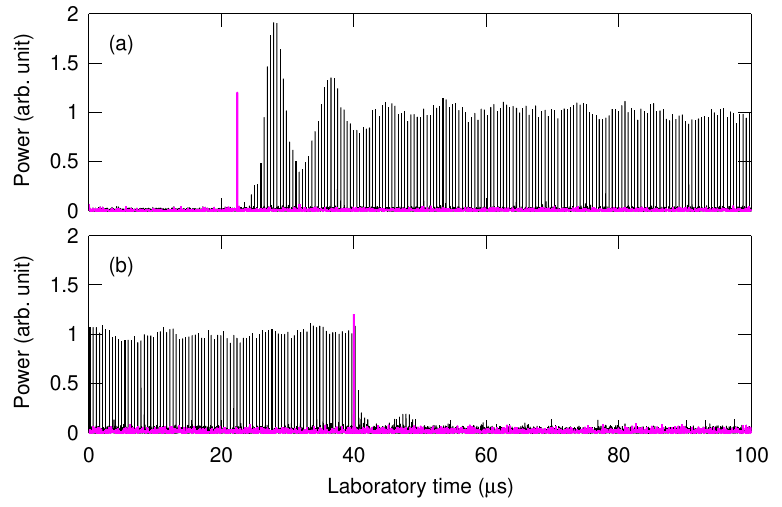}
	\caption{\small Experimentally measured dynamics of (a) CS writing and (b) erasing using intensity modulation. Black curves correspond to optical signals measured at the cavity output while magenta curves correspond to the addressing pulse.}
	\label{fig4}
\vskip-10pt
\end{figure}

In Fig. \ref{fig4}(a), we show experimentally measured CS writing dynamics. At $t = 22~\mu\mathrm{s}$, a single intensity modulation pulse with 28~W measured peak power is launched into the cavity (magenta curve). Within a couple of roundtrips, we see the emergence of a signal (black curve) at the cavity output whose 0.5~$\mu\mathrm{s}$ periodicity matches with the cavity round trip time. Because this signal is measured after the off-set band-pass filter, we can conclude that it represents a (picosecond-scale) short temporal feature circulating in the cavity. The feature shapes into a temporal CS in about 40~$\mu$s (i.e., 80 round trips), as evidenced by the stabilization of the signal amplitude. Next, in Fig.~\ref{fig4}(b) we show results that demonstrate CS erasure. Using the exact same parameters as for writing, we first excite a CS and allow it to stabilize. Then, at $t \approx 40~\mu\mathrm{s}$, the intensity modulation pulse is injected into the cavity, where it temporally overlaps with the CS thanks to the synchronization of the clocks driving the phase and intensity modulators. As can be seen, the CS [black lines in Fig. \ref{fig4}(b)] immediately attenuates and dies off in about 10~$\mu\mathrm{s}$  (i.e., 20 round trips). These results clearly demonstrate that, via direct intensity modulation of the cavity driving field, CS writing and erasing can be efficiently achieved using a single temporally broad addressing pulse.

\begin{figure}[t]
	\centering
	\includegraphics[width=\linewidth, clip = true]{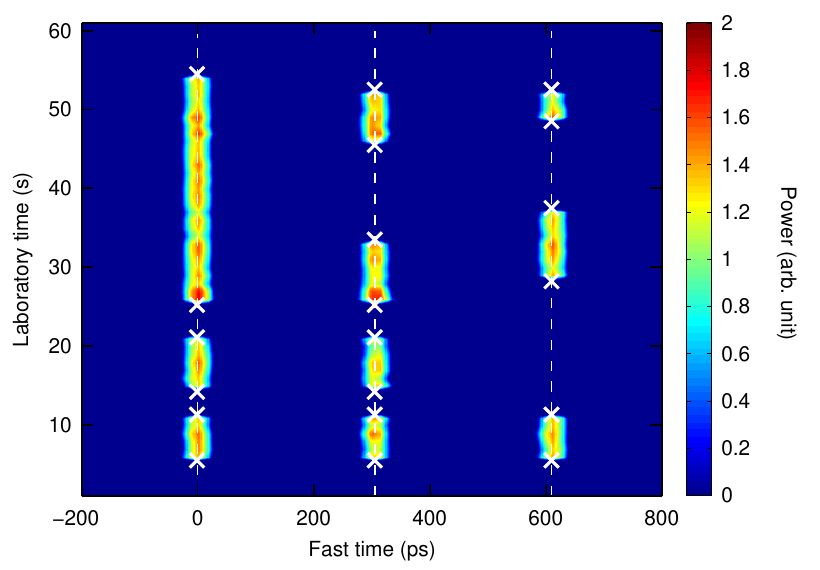}
	\caption{\small The manipulation of a 3-bit CS pattern via intensity modulation writing and erasing. The three dashed white lines indicate the temporal positions of the three phase modulation pulses. The white crosses highlight the instances when addressing pulses are applied to write or erase CSs. PD, photodetector.}
	\label{fig5}
\vskip-10pt
\end{figure}

Next, we demonstrate the flexibility of our technique by manipulating a simple binary buffer~\cite{jang_all-optical_2016}, where the presence or absence of a temporal CS represents a logical 1 or 0, respectively. To realise appropriate bit slots, we phase modulate the cavity driving field with a pattern consisting of three 60~ps pulses spaced by 300~ps. The false color plot in Fig.~\ref{fig5} shows the measured cavity output as the CS sequence is manipulated over one minute. The experiment starts from an empty buffer (i.e., with no CSs). At $t = 5~\mathrm{s}$, we launch a sequence of three intensity modulation pulses to simultaneously write three CSs spaced by 300~ps. These CSs persist until $t = 10~\mathrm{s}$, when we erase all of them by launching another set of three intensity modulation pulses into the cavity. Subsequently, we demonstrate more complex writing and erasing combinations by changing the sequence of intensity modulation pulses used for addressing. The fidelity throughout the demonstration was 100\% (i.e., there are no failed writing or erasing attempts), and we emphasize that (i) writing and erasing was achieved using the exact same addressing pulses and (ii) each individual addressing only required a single pulse launched into the cavity.

Before closing, we briefly comment on discrepancies  between our experimental findings and numerical simulations. First, the addressing pulse parameters needed for writing and erasing can differ considerably between our experiments and simulations: for example, 28~W peak power pulses result in the generation of a single CS in our experiments, whereas several would emerge in our simulations [using other parameters as in Fig.~\ref{fig1}]. Moreover, in our experiments, writing and erasure can be achieved without modifying any experimental parameters, whereas in numerical simulations, different parameters (e.g. addressing pulse peak power) are generally required. Second, CS writing occurs more rapidly in our experiments compared to our simulations: simulations realistically modelling the experiments show dynamics similar to that in Fig.~\ref{fig4}(a), but with a delay of several round trips. Third, our experiments show that, during CS erasure, the signal detected after the off-set band-pass filter decays immediately [c.f. Fig.~\ref{fig4}(b)], while numerical simulations predict the signal to first increase and only afterwards to decay. We believe these discrepancies arise partly due to uncertainties in experimental parameters (e.g. detuning): the addressing dynamics depend sensitively on the precise parameters used, and so small uncertainties can give rise to significant deviations. We also suspect that a major source of discrepancy lies in our crude modelling of the addressing pulse. That pulse undergoes both electronic and optical amplification, and propagates through an optical fiber before it is launched into the cavity. Accordingly, its intensity and phase profiles are likely to deviate from the simple Gaussian pulse used in our simulations. Further research is needed to obtain full quantitative (instead of qualitative) agreement between simulated and experimentally observed addressing dynamics.

In summary, we have demonstrated that a single temporally broad intensity modulation pulse can be used for efficient writing and erasure of temporal Kerr CSs. Moreover, we have presented a simple physical interpretation of the writing dynamics, and expect this interpretation to also be of relevance to proposed microresonator systems where CS writing is achieved by means of broad excitation pulses~\cite{kang_deterministic_2017}. Thus, in addition to enhancing our ability to address CSs in macroscopic fiber resonator, the results of our research could facilitate the systematic excitation of CSs in microresonators.

We acknowledge support from the Marsden Fund of the Royal Society of New Zealand. M. Erkintalo further acknowledges support from the the Rutherford Discovery Fellowships of the Royal Society of New Zealand. F. Leo acknowledges financial support from the European Research Council and the Fonds de la Recherche Scientifique (FNRS).



\end{document}